\documentstyle[twocolumn,aps,prl]{revtex}

\draft

\begin{document}

\twocolumn[\hsize\textwidth\columnwidth\hsize\csname
@twocolumnfalse\endcsname

\title{Constraints on the Interaction between Dark Matter and 
Baryons from Cooling Flow Clusters}

\author{Bo Qin and Xiang-Ping Wu}

\address{National Astronomical 
Observatories, Chinese Academy of Sciences, Beijing 100012, China}

\date{Accepted by {\it Phys. Rev. Lett.}}

\maketitle
\begin{abstract}
Other nongravitational heating processes are needed to resolve the 
disagreement between the absence of cool gas components
in the centers of galaxy clusters revealed recently by Chandra and XMM 
observations and the expectations of conventional radiative cooling models.
Here we propose that the interaction between dark matter 
particles and ordinary baryonic matter may act as an alternative for 
the reheating of intracluster medium (ICM) in the inner regions of clusters,
in which kinetic energy of dark matter is transported  
to ICM to balance the radiative cooling.
Using the Chandra and XMM data of typical clusters, we set a useful
constraint on the dark matter-baryon cross-section:
$\sigma_{xp}/m_x \sim 1 \times 10^{-25}$ cm$^2$GeV$^{-1}$, 
where $m_x$ is the mass of dark matter particles. 
\end{abstract}

\pacs{PACS numbers: 95.35.+d, 98.65.Hb, 98.65.Cw, 98.80.-k } 
]

The standard view of the cold dark matter (CDM) assumes that
dark matter consists of massive particles with weak interactions with 
ordinary baryonic matter, as well as with weak self-interactions. 
This scenario has proved to be very
successful in explaining the origin and evolution of 
cosmic structures  on large scales, but may conflict
with astrophysical observations on galactic and subgalactic scales. 
Many attempts have thus been made in recent years to modify
the standard CDM model. For example, the CDM particles may be
self-interacting \cite{Spergel00}, or the dark matter particles
could be warm \cite{Bode00}.

While it is widely believed that the interaction
of dark matter with ordinary matter is weak, 
the alternative picture, in which dark matter
interacts strongly with ordinary matter, still remains an
interesting issue. Indeed, a decade ago, Starkman 
{\it et al.} \cite{Starkman90} 
surveyed all current observations and experiments and found that 
there exist several allowed regions 
for the dark matter consisting of strongly interacting 
massive particles (SIMPs) which have large cross-sections
with ordinary matter.

Recently Wandelt {\it et al.} \cite{Wandelt00} 
have re-evaluated the issue and
presented a new constraint on SIMPs. 
Adding in the results from new experiments, 
they found that their stringent constraint allows a region in which
$\sigma_{xp}/m_x \sim 10^{-26}-10^{-24}$ cm$^2$GeV$^{-1}$, 
with $m_x > 10^5$GeV, 
where $\sigma_{xp}$ is the dark matter-proton cross-section, 
$m_x$ and $m_p$ are the masses of dark matter particles and  
protons, respectively.
Interestingly, this cross-section is similar to both the ordinary 
hadron-hadron cross-section $\sigma_{pp}/m_p$,
and the dark matter-dark matter cross-section
$\sigma_{xx}/m_x=8\times10^{-25}-1\times10^{-23}$ cm$^2$GeV$^{-1}$
in the self-interacting CDM scenario
proposed by Spergel and Steinhardt \cite{Spergel00}.
In fact, it was the striking similarity between 
$\sigma_{xx}/m_x$ and $\sigma_{pp}/m_p$
that had led Wandelt {\it et al.} to address
the possibility that dark matter particles may interact 
strongly with ordinary baryonic matter as well as with themselves.
It was argued that if the dark matter consists of an exotic, 
neutral and stable hadron which does not radiate photons
or other massless mesons, then the hadron will act like 
dissipationless, collisional dark matter.

On the other hand, 
recent high resolution observations of the hot X-ray emitting 
intra-cluster medium (ICM)
in cooling flow galaxy clusters with the new generation X-ray 
satellites Chandra 
\cite{David00} 
and XMM 
\cite{Arnaud01} \cite{Peterson01}
bring new puzzle to astrophysicists. 
It has long been realized that the 
radiative cooling, due to thermal bremsstrahlung, in the central 
regions of cooling-flow clusters is normally a rapid process, 
compared with the dynamical time  of the system. 
For typical rich clusters, the central cooling time is less than 
$10^9$~yr, an order of magnitude shorter than 
the Hubble time (see, e.g. Refs.
\cite{Sarazin} \cite{Fabian94}).
Hence, the ICM near the cluster centers should be found 
to be considerably cooler than that of the outer regions.

With unprecedentedly high spatial resolution and sensitivity, 
the Chandra/XMM observations
have revealed detailed temperature structures down to 
a few kpc from the centers of nearby clusters. 
Although a radial temperature gradient is clearly observed in each cluster,
the cooled gas component with temperatures lower than 
$\sim 2$ keV is prominently absent and 
the central ICM temperature is only found to be slightly lower 
than that of the outer regions. This seems to challenge conventional
theoretical expectations.

Various possibilities have been suggested to 
explain this discrepancy \cite{Peterson01}, which 
can be essentially classified into three categories.
1. Some additional heating mechanism, although still very uncertain, 
is needed to provide extra energy to the 
ICM in cluster centers and hence
to temper the cooling process, quickly reheating the cool 
gas back to the hot phase.
2. The gas has rapidly cooled below $\sim 2$~keV and we therefore
do not see it. 
3. The cool gas may exist, but its emission is absorbed by cold 
material at the center of the flow. 
Further X-ray observations are thus required to distinguish
these possibilities.

Yet, an alternative scenario is that the above inconsistency with the 
conventional cluster cooling picture may be alleviated if dark matter
interacts strongly with ordinary baryonic matter. 
This is because dark matter may serve as a heating source to 
transport its kinetic energy to ICM. Therefore, the problem
we will address is twofold: First, it provides a natural resolution
to the absence of the cool ICM components below $\sim 2$ keV 
in cluster centers.
Second, study of the distributions of the ICM density and temperature 
in clusters may allow us to set useful constraints 
on the dark matter-baryon cross-section $\sigma_{xp}$.

In galaxy clusters, dark matter particles
and  protons have similar rms velocities as they
trace the same underlying gravitational potential. 
However,  the ``temperature'' of the dark matter particles
should be much higher than that of the protons within the
framework of the standard CDM model. This simply arises from the
fact that CDM particles are much heavier than protons.

Now we consider that the protons and dark matter particles 
have Maxwellian velocity distributions with velocity 
dispersions $\sigma_p$ and $\sigma_x$, respectively. 
For a proton and a dark matter particle having velocities ${\bf u}$ 
and ${\bf v}$ in the lab frame and colliding at relative velocity 
${\bf s} = {\bf u}- {\bf v}$, the proton's outgoing speed 
in the center of mass (CM) frame will be $s$. Averaged over
all the directions, the proton's kinetic energy in the lab 
frame will be $m_p(s^2+v^2)/2$ which gives the averged energy
gain of the proton to be $m_p(s^2+v^2-u^2)/2$. Therefore the 
energy transfer rate from dark matter to protons is 
\begin{eqnarray}
\frac{dE}{dt\,dV} & = & 
	 \frac{m_p n_p n_x \sigma_{xp}} 
	      {4(2\pi)^3 \sigma_x^3 \sigma_p^3}
	 \int du \, 4\pi u^2 e^{-u^2/2\sigma_p^2}
	\nonumber \\
	&  \times &
	 \int dv \, 4\pi v^2 e^{-v^2/2\sigma_x^2}
	 \int_{-1}^{1} d\cos\theta \, (s^2+v^2-u^2) s, 
\end{eqnarray}
where $\cos\theta \equiv {\bf u}\cdot{\bf v}/uv$, and
$n_p$ and $n_x$ are the number densities of protons and
dark matter particles, respectively. It follows that  
the temperature change of the protons due to the heating 
of dark matter particles is
\begin{equation}
\frac{d\ln T}{dt} =  
	 \frac{n_x \sigma_{xp}} 
	      {3\pi \sigma_x^3 \sigma_p^5}
	 \int du \, u e^{-u^2/2\sigma_p^2} 
	 \int dv \, v e^{-v^2/2\sigma_x^2} f(u,v), 
\end{equation}
where 
\begin{eqnarray}
f(u,v) & = & 
	  \frac{(u+v)^5-\mid u-v \mid^5} {5} 
	 \nonumber \\
       & &
	  + (v^2-u^2) \frac{(u+v)^3-\mid{u-v}\mid^3}{3}.
\end{eqnarray}
Experiencing the same graviational field, the protons and dark 
matter particles in a galaxy cluster should have similar 
velocity dispersions. The place that one needs to be careful 
is the cluster center where the radiative cooling is the most 
rapid. However, as discovered recently by the Chandra and XMM
observations, the cooling process in the central regions of 
galaxy clusters is probably very inefficient and much slower 
than theoretical expectations (see below for further 
explanations). Therefore, it appears a reasonable approximation to 
assume that the protons and dark matter have the same velocity 
dispersion.  Thus, for simplicity, we take 
$\sigma_p = \sigma_x = \sigma$ in our evaluation. Integrating the 
right-hand side of Eq.(2), we finally obtain the equation describing
the heating process of protons by dark matter:
\begin{equation}
dT_{\rm heating} = \frac{25}{2\sqrt{\pi}} n_x \sigma_{xp} \sigma T dt
\end{equation}

Recall the definition of the cooling time 
$t_c \equiv - d\ln T /dt$ \cite{Sarazin}, 
which gives 
$dT_{\rm cooling} = -T \, dt/t_c$.
Now the overall cooling process of clusters is described by
\begin{equation}
dT = -T \, dt/t_c',
\end{equation} 
where 
\begin{equation}
\frac{1}{t_c'} = \frac{1}{t_c} 
		- \frac{25}{2\sqrt{\pi}} n_x \sigma_{xp} \sigma.
\end{equation}
From the cluster temperature profiles observed by Chandra and XMM,
the central temperatures of ICM is approximately half of 
the global values, 
which indicates that the cooling in cluster centers has been 
greatly tempered, and that $t_c'$, the quantity characterizing  the 
``actual'' cooling process, should be comparable to the cluster
age of $\sim 10^{10}$~yr. On the other hand, by measuring the electron 
density and temperature, Chandra and XMM have already obtained 
detailed information about $t_c$. For the central regions of radii 
$r \sim 100$~kpc, $t_c$ is only a few Gyr. And for the innermost 
regions of a few kpc, $t_c$ reaches its lowest value of about 
$3 \times 10^8$~yr.
This yields that $t_c' \gg t_c$, and hence we can set a constraint on 
the dark matter-baryon cross-section 
\begin{equation}
\frac{\sigma_{xp}}{m_x} = \frac{2\sqrt{\pi}}{25}
			  \frac{1}{\rho_x \,  \sigma \,  t_c}, 
\end{equation}
where $\rho_x = m_x n_x$ is the dark matter mass density.

From strong gravitational lensing observations, Tyson {\it et al.}
\cite{Tyson98}
obtained a detailed mass map in the central region 
of the cluster CL0024+1654, which does not 
contain a central cD galaxy, and found the existence of a 
$35h^{-1}$~kpc soft core 
(where $h\equiv H_0/100$, and $H_0$ is the Hubble constant
in units of km\,s$^{-1}$\,Mpc$^{-1}$). 
In particular, Firmani {\it et al.} 
\cite{Firmani}
presented an analysis of the halo mass for dwarf galaxies, low
surface brightness galaxies, and clusters of galaxies, and 
concluded that the halo central density is nearly independent
of the total mass over a broad mass range with an 
average value of $\sim 0.02M_{\odot}/pc^3$. 
Taking this characteristic value, together with the typical values of 
X-ray clusters, we have
\begin{eqnarray}
\frac{\sigma_{xp}}{m_x} & = & 6 \times 10^{-26}\,{\rm cm}^2\,{\rm GeV}^{-1}
	\left( \frac{\rho_x} {0.02M_{\odot}\,{\rm pc}^{-3}} \right)^{-1}
	\nonumber \\
	& &
	\times
	\left( \frac{\sigma} {1000\,{\rm km\,s}^{-1}} \right)^{-1}
	\left( \frac{t_c} {10^9 \,{\rm yr}} \right)^{-1}.
\end{eqnarray}

Note that 
$\rho_x$, $\sigma$ and $t_c$ are not independent of each other. 
The right-hand side of Eq.(7) can be further simplified by 
the employment of the cooling time estimate in terms of 
thermal bremsstrahlung \cite{Sarazin}
$
t_c = 1 \times 10^{9} {\rm yr} \, (n_e/0.1{\rm cm}^{-3})^{-1} 
        (T/10^8{\rm K})^{1/2} 
$,
where $n_e$ is the electron density which can be derived from 
X-ray observations. Alternatively, we introduce the local
baryon fraction $\tilde{f}_b$ such that 
$\rho_x = \tilde{f}_b^{-1} \, \mu_e m_p n_e$,
where $\mu_e=1.13$.  
Moreover, we replace the proton velocity dispersion $\sigma$ by 
the ICM temperature $T$ through 
$m_p \sigma^2 \approx kT$, where $k$ is the Boltzmann constant.
Finally, the cross-section $\sigma_{xp}$ is written as
\begin{equation}
\frac{\sigma_{xp}}{m_x}  =  9 \times 10^{-26}\,{\rm cm}^2\,{\rm GeV}^{-1}
	\left( \frac{\tilde{f}_b}{0.1} \right)
	\left( \frac{T}{5\times 10^7\,{\rm K}} \right)^{-1}.
\end{equation}
The radial dependence of the quantity $\tilde{f}_b(r)$ 
resembles approximately that of the volume-averaged 
baryon fraction $f_b$. The latter appears 
to be a slowly increasing function of radius
(see, e.g. Ref. \cite{Wu} and references therein), similar to
the radial variation of the ICM temperature $T$ as 
indicated by Chandra and XMM measurements. 
As a result, the overall effect of $\tilde{f}_b$ and $T$ 
makes the right-hand side of Eq.(9) roughly constant, or
insensitive to cluster radius.

For typical clusters, it is appropiate to choose 
$T=5\times 10^7$~K and $\tilde{f}_b=0.1$ in Eq.(9),
or $\sigma=1000$~km\,s$^{-1}$ and $t_c=5\times 10^8$~yr
in Eq.(8) for cluster central regions, which gives 
$\sigma_{xp}/m_x \sim 1\times 10^{-25}$ cm$^2$GeV$^{-1}$.
The $\sigma_{xp}/m_x$ value we obtained for the dark matter-baryon 
interaction is in good agreement with  
Wandelt {\it et al.}'s findings  \cite{Wandelt00} 
in their exclusion plot.
Moreover, it is similar to the $\sigma_{pp}/m_p$ value for the 
ordinary hadron-hadron strong interaction, and also near the lower 
end of the $\sigma_{xx}/m_x$ value for the dark matter-dark matter 
interaction in the self-interacting CDM scenario \cite{Spergel00},

Clearly, our $\sigma_{xp}/m_x$ value is suggestive of a strong 
interaction between dark matter and ordinary baryonic
matter, which challenges the conventional views towards dark
matter that the dark matter particles interact 
weakly with ordinary matter. Detailed theoretical and experimental 
discussions and explanations on SIMPs have been presented 
in full length by Starkman {\it et al.} \cite{Starkman90} 
and Wandelt {\it et al.} \cite{Wandelt00}. 
In this paper, we have no intention to enter into
the exploration of the particle physics of SIMPs. 
Rather, we would like to give 
an empirical evaluation of SIMPs from 
recent astronomical observations of cooling flow clusters.

To summarize, in this paper we have proposed that the puzzle with 
the cooling picture of galaxy clusters, arised from recent
Chandra and XMM high resolution observations, could be naturally 
settled if we assume that dark matter interacts strongly 
with ordinary matter. This interaction provides a
new heating mechanism for the baryons in some
gravitationally bound dynamical systems like galaxy clusters. 
Using the typical properties of density and temperature of ICM
revealed by Chandra and XMM, 
we have set a useful constraint on the 
dark matter-baryon cross-section 
$\sigma_{xp}/m_x \sim 1\times 10^{-25}$ cm$^2$GeV$^{-1}$. 
This estimate is  consistent with earlier findings of Refs. 
\cite{Starkman90} and \cite{Wandelt00} from a combined analysis of 
the existing experimental/observational constraints.

\section*{Acknowledgements}
We thank Zugan Deng for stimulating discussions. 
We are grateful to an anonymous referee for insightful comments 
which greatly improved the manuscript. This work was supported by 
the National Science Foundation of China under grants 1972531 
and 10003002, and the Ministry of Science and Technology of China, 
under grant No. NKBRSF G19990754.


\begin{thebibliography}{99}

\bibitem{Spergel00} D.N. Spergel and P.J. Steinhardt, Phys. Rev. Lett., 
84, 3760 (2000).

\bibitem{Bode00}
P. Bode, J.P. Ostriker, and Turok, N.,
astro-ph/0010389.

\bibitem{Starkman90} G. Starkman, A. Gould, R. Esmailzadeh, and S. Dimopoulos,
{\it Phys. Rev. D}{\bf 41}, 3594 (1990).

\bibitem{Wandelt00} B.D. Wandelt, R. Dave, G.R. Farrar, P.C. McGuire,
D.N. Spergel, and P.J. Steinhardt, astro-ph/0006344.

\bibitem{David00} L.P. David {\it et al}., Astrophys. J.
(in press, astro-ph/0010224);
S.W. Allen, S. Ettori and A.C. Fabian, 
Mon. Not. R. Astron. Soc.
(in press, astro-ph/0008517);
A.C. Fabian {\it et al}., 
Mon. Not. R. Astron. Soc.
(in press, astro-ph/0011547);
S.W. Allen  {\it et al}.,
Mon. Not. R. Astron. Soc.
(in press, astro-ph/0101162).

\bibitem{Arnaud01}M. Arnaud {\it et al}., 
Astron. Astrophys., 365, L80 (2001);
T. Tamura {\it et al}., 
Astron. Astrophys., 365, L87 (2001);
J.S. Kaastra {\it et al}., 
Astron. Astrophys., 365, L99 (2001).

\bibitem{Peterson01} J.R. Peterson {\it et al}., 
Astron. Astrophys., 365, L104 (2001).

\bibitem{Sarazin} C.L. Sarazin, X-ray Emission from Clusters
of Galaxies (Cambridge Univ. Press, 1988).

\bibitem{Fabian94} A.C. Fabian, 
Annu. Rev. Astron. Astrophys. {\bf 32}, 277 (1994);
D.A. White, C. Jones and W. Forman, 
Mon. Not. R. Astron. Soc., {\bf 292}, 419 (1997).

\bibitem{Tyson98} J.A. Tyson, G.P.  Kochanski, and I.P. Dell'antonio, 
Astrophys. J., {\bf 498}, L107  (1998).

\bibitem{Firmani} C. Firmani {\it et al}., 
Mon. Not. R. Astron. Soc., {\bf 315}, L29 (2000).

\bibitem{Wu} X.P. Wu and Y.J. Xue, 
Mon. Not. R. Astron. Soc., {\bf 311}, 825 (2000).


\end{thebibliography}
\end{document}